\documentclass[manuscript,screen]{acmart}

\usepackage{booktabs} 
\usepackage{amsmath}
\usepackage{array}
\usepackage{graphicx}
\usepackage{longtable}
\usepackage{diagbox}
\usepackage{footnote}
\makesavenoteenv{tabular}
\makesavenoteenv{table}
\usepackage{amssymb}
\usepackage{subfig}
\usepackage{mathtools}
\usepackage{soul}
\usepackage[font=small,skip=0pt]{caption}
\usepackage{array}
\usepackage{amsthm,enumitem}

\AtBeginDocument{%
  \providecommand\BibTeX{{%
    \normalfont B\kern-0.5em{\scshape i\kern-0.25em b}\kern-0.8em\TeX}}}

\setcopyright{acmcopyright}
\copyrightyear{2018}
\acmYear{2018}
\acmDOI{10.1145/1122445.1122456}

\begin{document}

\title{A First Look at References from the Dark to Surface Web World}

\author{Mahdieh Zabihimayvan}
\email{zabihimayvan.2@wright.edu}
\author{Derek Doran}
\email{derek.doran@wright.edu}
\affiliation{%
  \institution{Wright State University}
  \city{Dayton}
  \state{Ohio}
}

\renewcommand{\shortauthors}{Zabihimayvan and Doran}

\begin{abstract}
Tor is one of the most well-known networks that protects the identity of both content providers and their clients against any tracking or tracing on the Internet. So far, most research attention has been focused on investigating the security and privacy concerns of Tor and characterizing the topic or hyperlink structure of its hidden services. 
However, there is still lack of knowledge about the information leakage attributed to the linking from Tor hidden services to the surface Web. This work addresses this gap by presenting a broad evaluation of the network of referencing from Tor to surface Web and investigates to what extent Tor hidden services are vulnerable against this type of information leakage. The analyses also consider how linking to surface websites can change the overall hyperlink structure of Tor hidden services. They also provide reports regarding the type of information and services provided by Tor domains. Results recover the dark-to-surface network as a single massive connected component where over 90\% of Tor hidden services have at least one link to the surface world despite their interest in being isolated from surface Web tracking. 
We identify that Tor directories have closest proximity to all other Web resources and significantly contribute to both communication and information dissemination through the network which emphasizes on the main application of Tor as information provider to the public. 
Our study is the product of crawling near 2 million pages from 23,145 onion seed addresses, over a three-month period.
\end{abstract}

\ccsdesc{Security and privacy~Security services}
\ccsdesc{Networks~Network properties}
\ccsdesc{Information systems~Deep web}

\keywords{Tor, Anonymity network, Information leakage, Dark Web}

\maketitle

\section{Introduction}

Anonymity networks prevent traffic analysis and network activity monitoring by complicating any possible tracking or tracing of users identity on the Web. The ``Dark Web'' defines a class of anonymity networks that require unique application layer protocols and authorization schemes to access the Internet. Tor~\cite{secondGeneration} has emerged as the most popular dark nets in contrast to many other dark networks such as I2P~\cite{cyberSecurity}, Riffle~\cite{riffle}, and Freenet\cite{freenet}. It provides anonymous communication for both senders and receivers using an encryption scheme similar to onion routing~\cite{goldschlag1999onion}. Tor hidden services, also called Tor domains, further provide privacy for users who run Internet services on Tor. Tor clients can directly access it using a Tor Web browser that provides their anonymity while browsing the Internet. 
In addition to the direct way, users can utilize Tor proxies such as \textit{tor2web} that alleviates connection with no need to install the Tor Web browser. In this way, client types the hostname of onion address in a regular Web browser and replace `.onion' with `.tor2web.io'. Then, tor2web, as a ``middleman'' between the client and the hidden server, retrieves the requested content from dark Web and forwards it to the client. 
However, if the requested page has any hyperlink to surface/regular websites, their addresses will not be rewritten by the Tor proxy. Hence, the Web browser used by the client can observe the identity of client and bypass the anonymization mechanism. 

Investigating the security and privacy threats on Tor has been attracted significant attention~\cite{shining}~\cite{trawling}~\cite{biryukov2015bitcoin}~\cite{cambiaso2019darknet}~\cite{bauer2007low}~\cite{sanatinia2017off}~\cite{sanatinia2019privacy}. Information leakage is a privacy threat that attributes to the connections onion domains have with the surface Web~\cite{onionEyes}. There are two third-parties that can take advantage of this information leakage: (1) destination domains on the \textbf{surface Web}. As client traffic can be monitored by the Web browser, the destination domain is able to observe the identity of the Tor user. A related work reports that Google, Facebook, and Twitter can respectively observe 13.20\%, 1.03\%, and 0.88\% of the traffic towards Tor hidden services~\cite{onionEyes}. (2) \textbf{A malicious hidden service} simply links to a surface website, both controlled by the same attacker, and monitors the traffic of users who visit this surface website using Tor proxies. The rise in the number of hyperlinks from dark domains to surface websites consequently increases the probability of information leakage in both scenarios. 

In spite of recent efforts to discover privacy concerns triggered by linking from Tor to surface Web~\cite{onionEyes} and to study the reference network of hidden services in Tor~\cite{bernaschi2019spiders}~\cite{burda2019characterizing}~\cite{griffith2017graph}~\cite{zabihimayvan2019broad}, there has been no study on characterizing the network of references from dark to surface Web. 
Such an investigation can reveal to what extent the Tor network is vulnerable against the information leakage caused by referring to surface websites and could give hints on how to improve the design of Tor hidden services to avoid such threats. 
To this end, we conduct a micro-level investigation on network of references from dark domains to surface websites using well-known network measurements and consider how the network generated by the dark-to-surface references differs with the linking pattern of Tor hidden services. We scrutinize results regarding the type of content and information provided by Tor services. 

Our findings reveal that, in contrast to the hyperlink structure of Tor domains, the dark-to-surface reference network is a small world where most resources participate in a single massive connected component with a small number of isolated Tor domains.
Investigation on neighbourhood diversity of Tor domains indicates only a few Tor hidden services are immune against information leakage caused by linking to surface Web while over 90\% contain at least one link to the surface. 
Moreover, in spite of the fact that dark domains intend to remain isolated, linking to surface raises their tendency to cluster. 
We also evaluate the types of services Tor domains provide and find that popular Tor directories have the maximum proximity to Web resources and significantly contribute to both communication and information dissemination through the dark-to-surface network. 
A few domains with other types of services including news, forums, pornography, and multimedia have also significant role in propagating information through the network. Analysis on the degree distribution shows 
popular surface websites such as GitHub, Twitter, and Facebook also receive large number of references from dark domains which indicate that they are popular even among Tor services.

This paper is organized as follows: Section~\ref{sec:rw} discusses related work on characterizing the Tor network as well as studies conducted on Tor security issues. Section~\ref{sec:dc} presents the description and basic statistics of data used in the analyses. Section~\ref{sec:linkP} investigates linking process of Tor services to dark/surface resources and presents the evaluation results. Section~\ref{sec:rview} analyzes the reference view of Tor services using well-known network metrics. Finally, Section~\ref{sec:con} summarizes the main conclusions and discusses the future work.
\section{Related Work}
\label{sec:rw}

Previous research on Tor can be categorized into two classes: (1) work which has focused on Tor relay network, understanding the security, privacy, and topological properties of Tor network; and (2) studies on characterizing different types of information and services hosted on Tor.

Towards understanding Tor security and privacy issues, 
Mohaisen {\em et al.} studied the possibility of observing Tor requests at global DNS infrastructure that could threaten the private location of servers hosting Tor services, and name/onion address of Tor domains~\cite{leakage}. 
McCoy {\em et al.} tried to answer how Tor is (mis-)used and what clients and routers contribute to this usage~\cite{shining}. They also proposed some remedies to improve the implementation of Tor network. 
Focusing on privacy of Tor hidden services, Biryukov {\em et al.} analyzed the traffic of services to evaluate their vulnerability against deanonymizing and take down attacks~\cite{trawling}. They demonstrated how current flaws in design and implementation of Tor hidden services can help attackers find the popularity of a hidden service, harvest its descriptor in a short time, and find its guard relay. They further proposed a large-scale attacking technique to disclose the IP address of notable number of Tor hidden services over one year. All the proposed techniques are evaluated over Silk Road, DuckDuckGo search engine, a case of a botnet that utilizes Tor hidden services as command/control channels.
Biryukov and Pustogarov indicated how using bitcoin over Tor can help man-in-the-middle attacks to fully observe information transmitted between Tor clients who use Bitcoin cryptocurrency~\cite{biryukov2015bitcoin}. 
In~\cite{cambiaso2019darknet}, Cambiaso {\em et al.} presented an exhaustive investigation on different types of security concerns over Tor network. Based on purpose of the attack, various threats are categorized into client-side, server-side, and network attacks. 
Bauer {\em et al.} investigated how flaws in Tor routing optimization approach can cause end-to-end traffic analysis attack~\cite{bauer2007low}. 
They evaluated the proposed attacks on PlanetLab and propose solutions to mitigate the severity of the threats. 
Sanatinia and Noubir demonstrated a man-in-the-middle attack on Tor hidden services that is mounted by an attacker who knows the private key of a hidden service~\cite{sanatinia2017off}. 
In another work, Sanatinia {\em et al.} investigated the longevity of Tor hidden services with respect to the privacy of Tor users and services~\cite{sanatinia2019privacy}. They indicated how it is possible to estimate the lifetime of hidden services with fairly high accuracy using a small percent of Tor HSDir relays. 

The topological properties of Tor, at physical and logical levels, are only beginning to be studied. Xu {\em et al.} quantitatively evaluated the structure of four terrorist and criminal related networks, one of which is from Tor~\cite{topology}. 
Sanchez-Rola {\em et al.} conducted a broader structural analysis over 7,257 Tor domains~\cite{onionEyes}. They find a surprising relation between Tor and the surface Web: there are more links from Tor domains to the surface Web than to other Tor domains. 
Bernaschi {\em et al.} presented a characterization study on topology of Tor network graph and investigated the persistence of hidden services and their hyperlinks~\cite{bernaschi2019spiders}. All analyses are conducted over three different snapshots of Tor captured during a five-month period. They also compared Tor with other social networks and surface Web graphs using well-known metrics. 
In another similar work~\cite{bernaschi2017exploring}, Bernaschi {\em et al.} investigated measurements to evaluate and characterize Tor hidden services data and topology of their network. They provided a critical discussion on possible data collection techniques for dark Web and conducted analyses on the relationship between Tor English content and its topology.
Focusing on deployment and mirroring of Tor hidden services, Burda {\em et al.} provided an extensive investigation on redundancy of Tor services across time and space~\cite{burda2019characterizing}.  
Similarly, Griffith {\em et al.} investigated the graph theoretic properties of Tor network and compared it with previous analyses conducted on the surface Web~\cite{griffith2017graph}. The study considers bow-tie structure, robustness and fragility against node removal, and the importance of reciprocal connections in the comparisons. 
\section{Dataset Description}
\label{sec:dc}

We conducted a comprehensive multi-threaded crawl of the Tor network to collect data for this research. 
The crawling strategy extracts the HTML of any Tor website reachable by a depth-first search up to depth 4 beginning from a seed list of 23,145 onion addresses. The seed list was built by merging the list used in a recent study~\cite{onionEyes} into those achieved by manual exploration of Reddit, Quora, Ahima, and other well-known surface Web directories during the time performing the crawls. 
Although a manually collected seed list may carry the risk of missing some parts of the dark network, its hidden nature makes it unlikely for there to ever be a single major directory of Tor Hidden services. We are confident that this way results in a broad crawling of Tor because Reddit, Quora, Ahima, and other surface Web directories are well-known among Tor users for providing the latest news and information about Tor and its hidden links. It suggests that these seeds are at worst practically helpful, and at best are ideal as entry points to the world of Tor. The crawlers are also assigned to cover all hyperlinks up to depth 4 from every entry page to make the data collection as broad as possible.

Due to rapid changes in content and structure of Tor~\cite{onionEyes} and temporary downtime for some domains, three crawls were repeated in June, August, and September 2019. To control the variability in the up and down time of hidden services, the union of the Tor sites collected during the three crawls were considered for the analysis. To prevent any access control polices, request rate limiters, and crawler blockers which may interrupt the data collection, crawlers only request HTML and follow hyperlinks, and do not download the full content of Web pages. A total of 1,905,714 distinct pages were captured across all crawls. To classify the crawled pages as being from the surface or from Tor, any webpage with suffix `.onion'  was classified as a Tor page, while the remainder are considered to be from the surface Web. 

\begin{figure}
\includegraphics[width=0.85\textwidth]{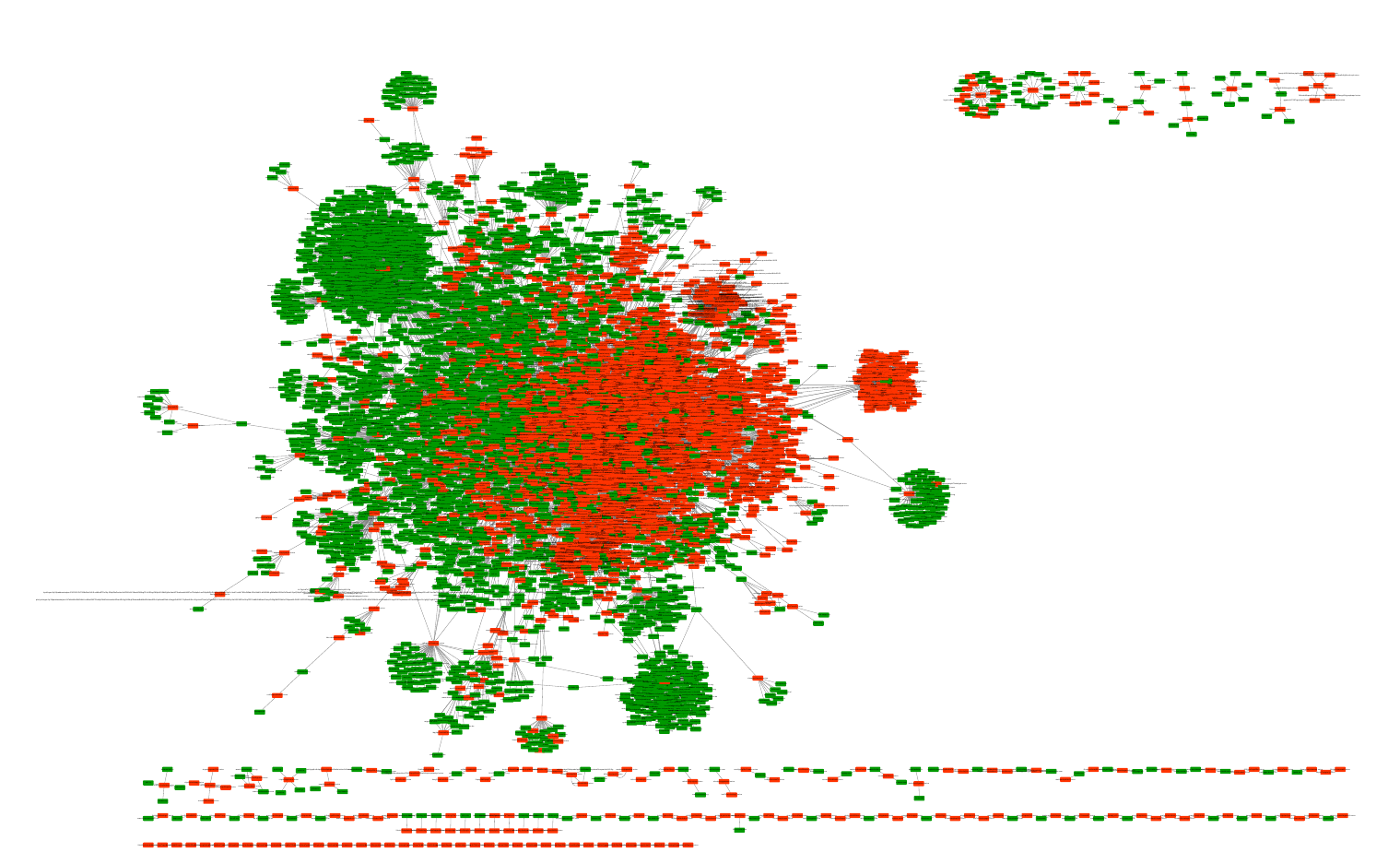}
\caption{Network of data collected during crawling}
\label{fig:net}
\end{figure}

We build a graph (presented in Figure~\ref{fig:net}) where vertices represent dark Web domains and surface websites and a directed relation indicates a hyperlink from a dark Web domain to a dark or surface Web domain. We avoided exploring the surface websites; thus, the graph does not contain any edges from surface Web nodes to other vertices. The investigation considers how Tor domains link to other resources on either dark or surface Web and presents analyses on the characteristics of this hyperlink structure conditioned by the type of service they provide. 

First, we examine some basic statistics of the network indicated in Table~\ref{tab:stat}. We find a sparse network with only 29,519 directed edges and 6,882 nodes~\footnote{All self-loops and multi edges have been removed from this graph.}. 
Nodes colored in red in Figure~\ref{fig:net} indicate Tor domains and their population (3,288 nodes) is approximately equal to the number of surface websites (3,593) which are green nodes. 
The average shortest path length also represents that by average, each node is reached from any other node using 4 edges.

The network also includes 127 weakly connected components where the largest one contains 6,564 vertices (95\%) and the others with 28 and 13 nodes are the second and third largest connected components of this network. This implies that the pattern Tor services use to link to surface/dark resources make almost all the nodes connected in one single massive component. 
Hence, similar to other socio-technical networks~\cite{connectedWeb}, the formation process of this network encourages linking to others to alleviate information dissemination through the network. 
The second and third connected components contains links between bitcoin and gambling Tor services with related websites on the surface Web which are not referred by other Tor domains. 
\begin{table}
\caption{Basic parameters of the network}
\begin{tabular} {| c | c |}
\toprule
Basic Parameter & Value\\
\midrule
No. Nodes ($|V|$) & 6882\\
No. Edges ($|E|$) & 29519 \\
No. Isolated nodes & 39\\
Density ($\rho$) & 0.0\\
No. Connected Components & 127 \\[-1pt]
Avg. Shortest Path length & 4.108\\
\bottomrule
\end{tabular}
\label{tab:stat}
\end{table}

To gain a perspective on the information provided by Tor hidden services, we employ an unsupervised content discovery and labeling procedure that we previously proposed~\cite{zabihimayvan2019broad}. The process runs the content of every Tor Web page~\footnote{where content is defined as any string outside of a markdown tag} through the Latent Dirichlet Allocation (LDA)~\cite{lda} and Graph-based Topic Labeling (GbTL)~\cite{gbtl} algorithms to derive a collection of semantic labels indicating broad topics of content on Tor. Each Tor hidden service is then assigned a label by the dominant topic present across the set of all Web pages investigated in the domain. 
Table~\ref{tab:summary} presents a summary description of the topic labels inferred.
\begin{table}
\caption{Summary description of Tor domain types}
\centering
\begin{tabular} { p{20mm} | p{130mm}  }
\toprule
Topic label & Description \\
\midrule
Directory & Directories include unnamed pages with address lists of Tor hidden services including TorDir and Hidden Wiki services, as well as search engines like DuckDuckGo and Fresh Onions.\\
\hline
Bitcoin & Provides services for Bitcoin transactions and fund transfers to wallets.\\
\hline
News & Host pages akin to personal weblogs where authors write essays on various topics and visitors can post follow-up comments.\\
\hline
Email & Provides communication services like email, chat room, and Tor VPNs.\\
\hline
Multimedia & Provides multimedia products like e-books, movies, music, games, and academic and press articles even if they are copyright protected.\\
\hline
Shopping & Shopping domains allow visitors to purchase goods and services, including drugs, medicine, as well as consultancy and investment services.\\
\hline
Forum & Forum domains host bulletin board and social network services for Tor users to discuss ideas and thoughts.\\
\hline
Gambling & Gambling domains offer services to bet money on games, to purchase gambling advice and consulting, and to read gambling-related news.\\
\hline
Dream market & A shopping domain so large that it merited its own topic category is the Dream market. It suggests a wide range of content available for sale, most of which are illicit.\\
\bottomrule
\end{tabular}
\label{tab:summary}
\end{table}
%
\section{Linking process of Tor services to dark/surface resources}
\label{sec:linkP}

\begin{figure}
\includegraphics[width=0.45\textwidth]{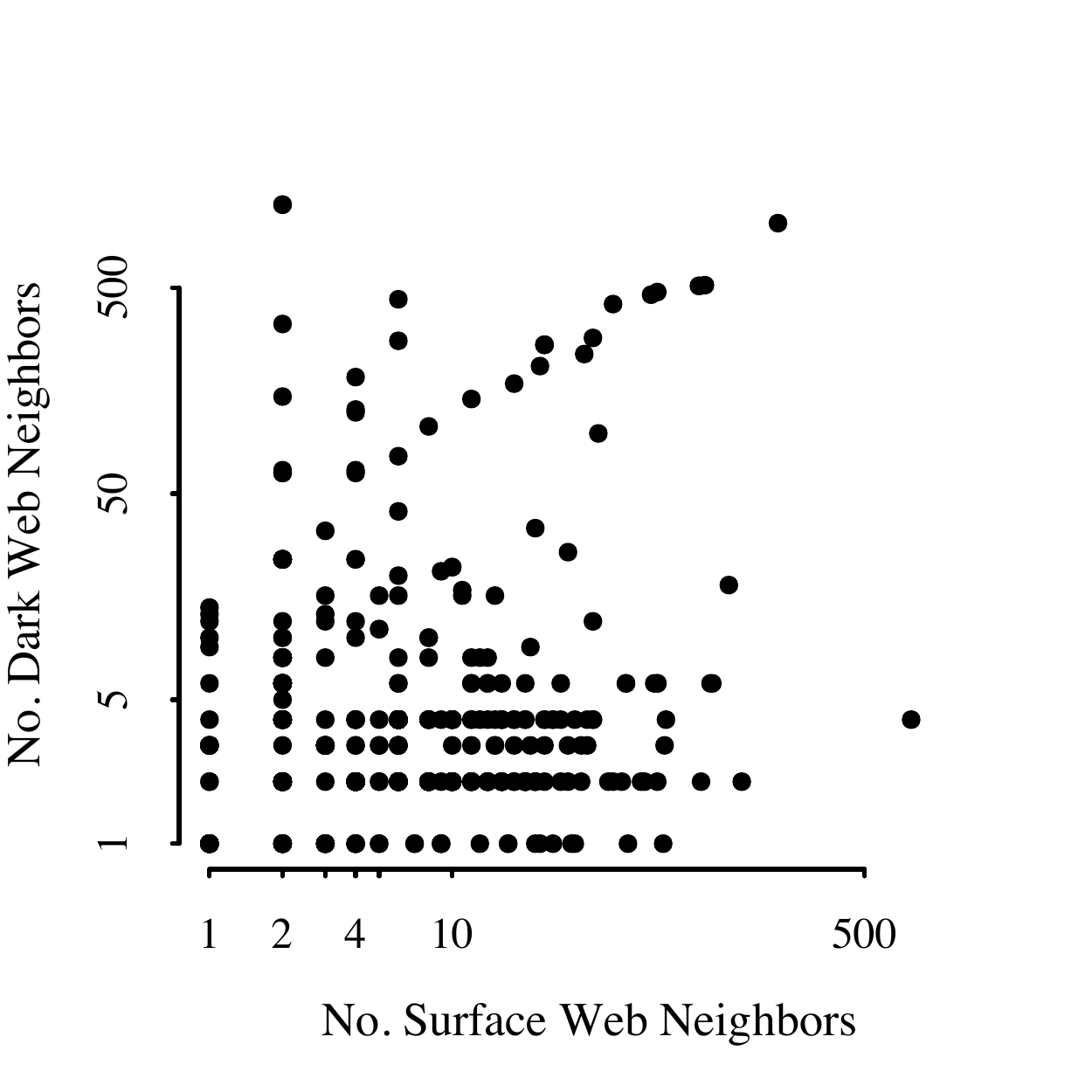}
\caption{Number of references to surface Vs. dark Web; both axes in logarithmic scale.}
\label{fig:scatter}
\end{figure}

\begin{figure}
\includegraphics[width=0.45\textwidth]{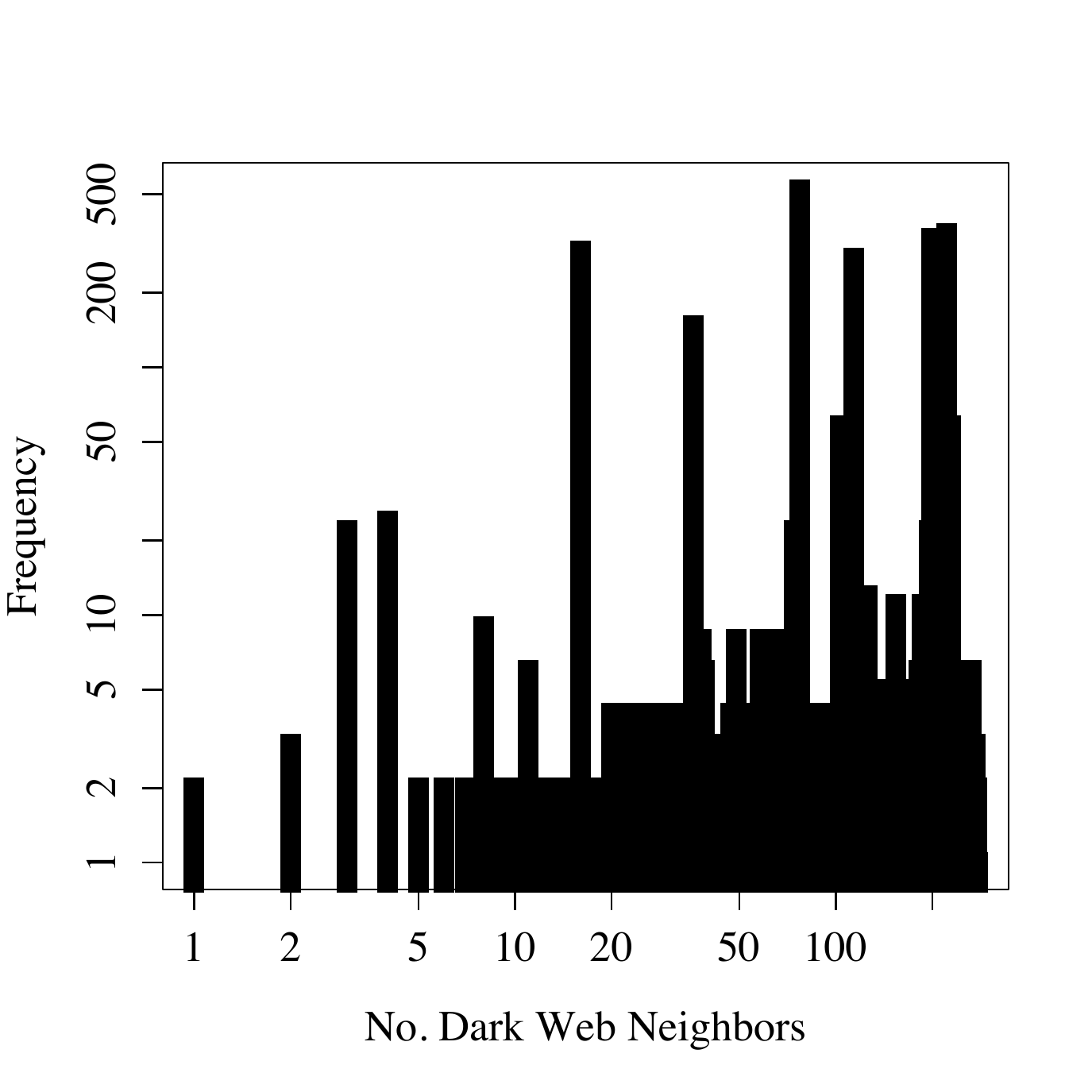}
\caption{Distribution of number of Dark Web neighbors for Tor domains in the first category; both axes in logarithmic scale.}
\label{fig:histDark}
\end{figure}

To gain an understanding of the linking process in the network, Figure~\ref{fig:scatter} indicates the number of hyperlinks from each Tor domain to Web pages on the surface and dark Web. From the vertical and horizontal lines in the scatterplot, points can be categorized into three divisions: 

(1) Services with less than 3 surface Web neighbors: 
These Tor services are comprised of 13\% of dark services while 8\% only have links to Tor domains with no reference to surface websites. Hence, only less than one tenth of dark services are immune against the information leakage caused by linking to the surface Web. 
Figure~\ref{fig:histDark} represents the distribution of reference sizes (number of references) for the first category in logarithmic scale. 
Values with the largest frequency indicate that most Tor domains in this class have around 50 to 100 hyperlinks to the other dark counterparts. 
Manual investigation on the domains with maximum number of dark neighbors shows that they belong to 31 Tor domains which is approximately equal to .94\% of all Tor services studied in this research. Exploring their corresponding services indicates that they are some well-known dark directories which present large lists of onion addresses in Tor network. For instance, one of the domains with maximum number of dark neighbors is a Tor directory (\textit{5jgis47vdcpaeafp.onion}) presenting 1,264 hyperlinks to other onion services in Spanish.
%
%

(2) Services with less than 3 dark Web neighbors: 
This category is comprised of 20\% of all Tor domains where 14\% are dark services which have links to the surface Web, but no reference to dark domains. This implies that less than one fifth of dark services contain hyperlinks to only surface websites. 
Table~\ref{tab:example} presents description of some Tor domains with number of surface Web neighbors close to the maximum (500). The domains listed are some of popular domains that provide services for digital cryptocurrency, hacking, and hosting Tor services. Users of such domains are referred to surface websites to perform bitcoin transactions, get latest news and information, or download multimedia resources such as electronic books.
\begin{table}
\caption{Description of some onion domains with more than 200 references to surface Web and no link to dark Web}
\centering
\begin{tabular} { c | c | c }
\toprule
Onion Domains & Service & \shortstack {\# of Surface \\ Web Neighbors}\\
\midrule
222222b63tfqitrx & Bitcoin & 188\\
222222zsr3ih57iw & Bitcoin & 196\\
3wcwjjnuvjyazeza & Hacking books & 697\\
4do6yq4iwstidagh & \shortstack {Hosting Debian \\ services} & 158\\
4sy6ebszykvcv2n6 & News (German) &1358\\
5nca3wxl33tzlzj5 & \shortstack {Hosting Debian \\ services} & 232\\
\bottomrule
\end{tabular}
\label{tab:example}
\end{table}

(3) The rest of points belong to dark services that contain hyperlinks to both surface and dark Web. This category forms 78\% of all points on the plot which in addition to the size of second category implies that more than 90\% of Tor domains contain at least one link to the surface Web. It indicates to what extent dark services provided on Tor can be vulnerable against information leakage due to linking to surface websites. 
This finding can also explain the reason of significantly smaller number of isolated nodes (1.1\% of Tor services) in the network of Figure~\ref{fig:net} in contrast to the Tor network studied in~\cite{zabihimayvan2019broad} where 12\% of crawled services were isolated.

As an example of services in this group, \textit{CensorBib}\footnote{3wcwjjnuvjyazeza.onion} is a Tor service linking to 921 dark domains and 205 websites from the surface Web. It is a well-known Tor service providing open access to research papers in the field of Internet censorship and Tor. For each article, there are three resources provided as hyperlinks: pdf, cached pdf, bib. The PDF version of each paper (if exists) redirects users to a publication website on the surface Web while the cached PDF version redirects users to an onion address to download the paper. The bib also contains the content of `.bib' file provided for each article. 
\section{Analyzing reference view of Tor services}
\label{sec:rview}

\begin{figure}
\includegraphics[width=0.85\textwidth]{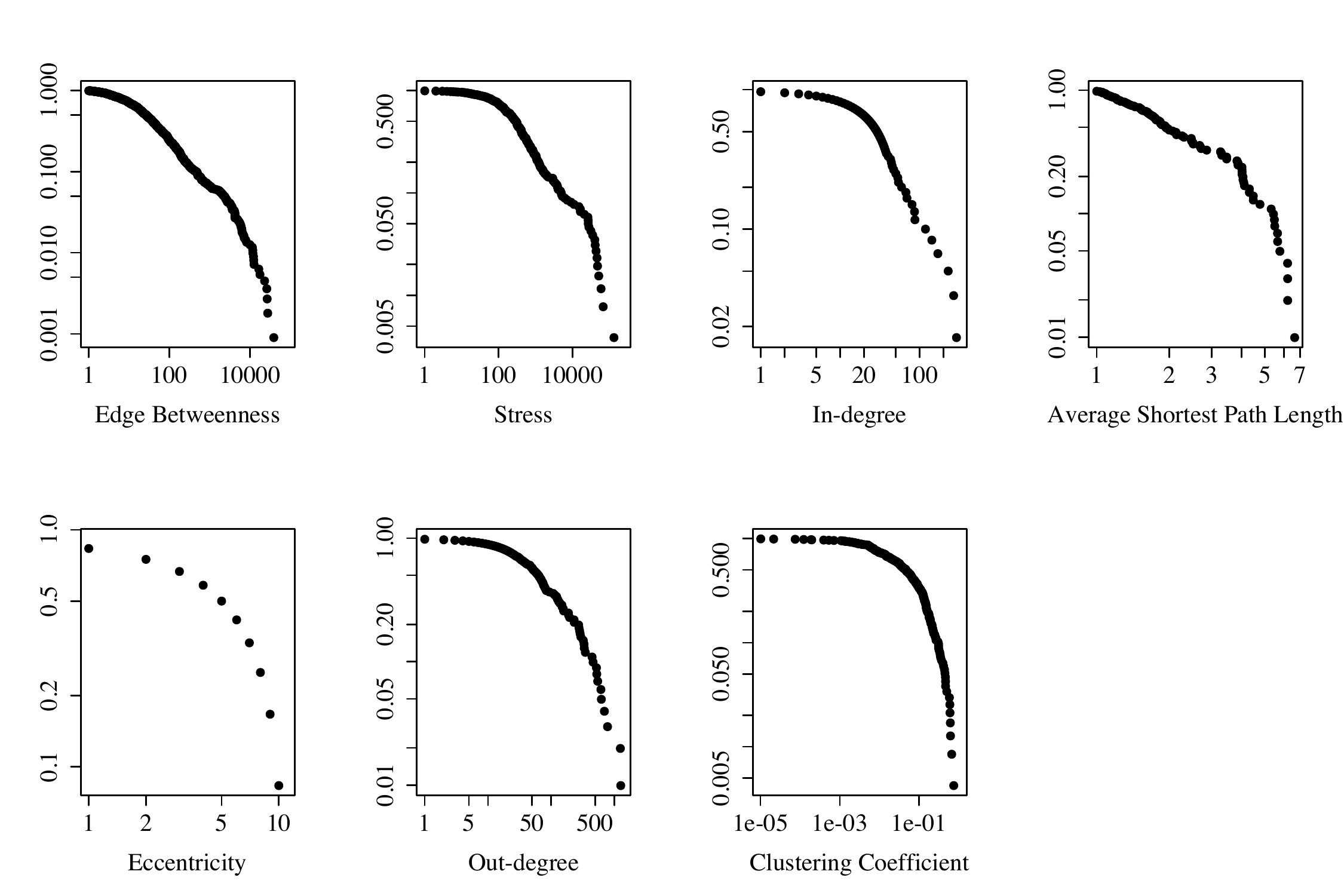}
\caption{Frequency Distribution of Network Parameters}
\label{fig:netPar}
\end{figure}

Now, we perform topological analysis on the directed network of hyperlinks from dark domains to other dark/surface resources using various measures of network properties. Table~\ref{tab:params} presents the basic statistics of network features. As illustrated, distributions of all network parameters are right-skewed since the mean is greater than the median, right tail of the distribution is long, and median is closer to the first quartile. 
Figure~\ref{fig:netPar} presents the CCDF plot of the distribution of each feature in log-log scale. CCDF, complementary cumulative distribution function, of a random variable $\nu$ shows the probability that $\nu$ will be greater than a particular value $x$~\cite{ccdf}.
\begin{table*}
\caption{Summary statistics of network features}
\centering
\begin{tabular} { c | c | c | c | c | c | c }
\toprule
Parameter & Min & $1$st. Qu & Median & Mean & $3$rd. Qu & Max \\
\midrule
Edge betweenness Cent. & 0.00 & 1.00 & 3.00 & 59.75 & 28.00 & 81,126.53 \\
\hline
Avg. Shortest Path Len. & 0.00 & 0.00 & 0.00 & 0.21 & 0.00 & 6.64 \\
\hline
Eccentricity & 0.00 & 0.00 & 0.00 & 0.25 & 0.00 & 11.00 \\
\hline
In-degree Cent. & 0.00 & 2.00 & 3.00 & 4.29 & 4.00 & 309.00 \\
\hline
Out-degree Cent. & 0.00 & 0.00 & 0.00 & 4.29 & 0.00 & 1,358.00\\
\hline
Stress Cent. & 0.00 & 0.00 & 0.00 & 175.00 & 0.00 & 230,857\\
\hline
Clustering Coef. & 0.00 & 0.00 & 0.00 & 0.09 & 0.083 & 1.00 \\
\bottomrule
\end{tabular}
\label{tab:params}
\end{table*}
%

\textbf{1. Edge betweenness centrality.}~\footnote{It is calculated for networks with no multiple edges.} The edge betweenness centrality of an edge $e$ indicates number of shortest paths between each pair of nodes that traverses through $e$ divided by total number of shortest paths between that pair of nodes~\cite{newman2004finding}. Assuming $Cent_{b}(e)$ as the edge betweenness centrality of edge $e$, the partial betweenness value of this edge for each pair of nodes is calculated as follows:
$$
b_{vt}(e)=\frac{\delta_{vt}(e)}{\delta_{vt}}
$$
where $\delta_{vt}(e)$ indicates number of shortest paths between $v$ and $t$ that contains the edge $e$ and  
$\delta_{vt}$ is the number of shortest paths between $v$ and $t$. $Cent_{b}(e)$ is calculated as the summation of the partial values for all pairs of vertices:
$$
Cent_{b}(e)=\sum_{v\neq t \in V}b_{vt}(e)
$$

A high value of this centrality for an edge implies that it plays an important role in bridging communication between network nodes. 
As mentioned before, the network does not contain any edges from surface Web nodes to other vertices. Therefore, edges here indicate connections between a pair of Tor services (dark-dark edges) or from one Tor domain to a surface website (dark-surface edges).
According to Table~\ref{tab:params}, average edge betweenness centrality is 59.75 which has significant difference with the maximum centrality (81,126.53). 
Figure~\ref{fig:netPar} indicates that the probability of values greater than 100 is less than 0.1 which has significant difference with probability of values lower than 100. 
Also, all dark-surface edges should have low edge betweenness centralities because as mentioned before, there is no link from surface websites to other vertices 
and thus, the edges that connect Tor to surface Web will have low edge betweenness centrality. 
On the other hand, investigation indicates that there is only 3\% of the edges with centralities greater than the average. This implies that few dark-dark edges have notable contribution to communication through the network and the most efficient nodes through the dark Web thus pass through the surface Web. 

To gain an overall impression of edges with high edge betweenness centrality, Table~\ref{tab:betwExamp} presents the top-six edges and type of services provided by their corresponding source and target domains. As illustrated, domains corresponding to edges with high edge betweenness centralities are well-known Tor information providers as either directory or news/forum services. This finding emphasizes the ability for information disseminating from the dark Web can be released over the world.
\begin{table}
\caption{List of edges with Edge betweenness Centralities greater than 10,000}
\centering
\begin{tabular} { c | c | p{35mm} | p{35mm} }
\toprule
Source & Target & Source service & Target service \\
\midrule
dirnxxdraygbifgc & wiki5kauuihowqi5 & Dir~\footnote{Dir stands for `Directory'} (OnionDir) & Dir (HiddenWiki)\\
\hline
doe6ypf2fcyznaq5 & dirnxxdraygbifgc & Dir (Runion Wiki~\footnote{Russian forum on the address of lwplxqzvmgu43uff.onion (for more details, please see https://darknetlive.com/runion-forum/)
}) & Dir (Web proxy address of OnionDir~\footnote{http://dirnxxdraygbifgc.onion/
}) \\
\hline
hiddenwik55b36km & newsiiwanaduqpre & Dir (HiddenWiki) & News (OnionNews providing news on Tor)\\
\hline
lwplxqzvmgu43uff & doe6ypf2fcyznaq5 & Forum (Runion Dark Forum) & Dir (Web proxy address of OnionDir)\\
\hline
deurfnquin7mvni2 & godnotabatovgyqz & Dir (Pyrowiki|Pyrotechnics drug wiki) & Dir (Fresh Onions search engine\\
\hline
newsiiwanaduqpre & dirnxxdraygbifgc & News (OnionNews providing news on Tor) & Dir (OnionDir)\\
\bottomrule
\end{tabular}
\label{tab:betwExamp}
\end{table}
%

\textbf{2. Eccentricity.} The eccentricity of a vertex is the inverse of maximum non-infinite distance it has from other vertices in the graph~\cite{hage1995eccentricity}. Distance between two nodes is the length of the shortest path between them where length of a path is declared as the number of edges building the path. Eccentricity of an isolated node is zero. By computing the shortest path between the vertex $v$ with others, we calculate the eccentricity of $v$, $Ecc(v)$, as:
$$
Ecc(v)=\frac{1}{dist(v,k)}
$$
where $k$ is the node which has longest shortest path with $v$ and function $dist$ returns the distance between them. High value of eccentricity implies that other vertices are in proximity of $v$ while low eccentricity indicates at least one node and its direct neighbors are located far from $v$. 
Maximum value of eccentricity can also depict the diameter of a network which is defined as the largest distance between two nodes. For a disconnected network, diameter is the maximum diameter of all the connected components. As Table~\ref{tab:params} demonstrates, the diameter of the network in Figure~\ref{fig:net} is equal to 11.0. The radius is also defined as the minimum none-zero eccentricity of a network which is here equal to 1.0. 
\begin{figure}
\includegraphics[width=0.45\textwidth]{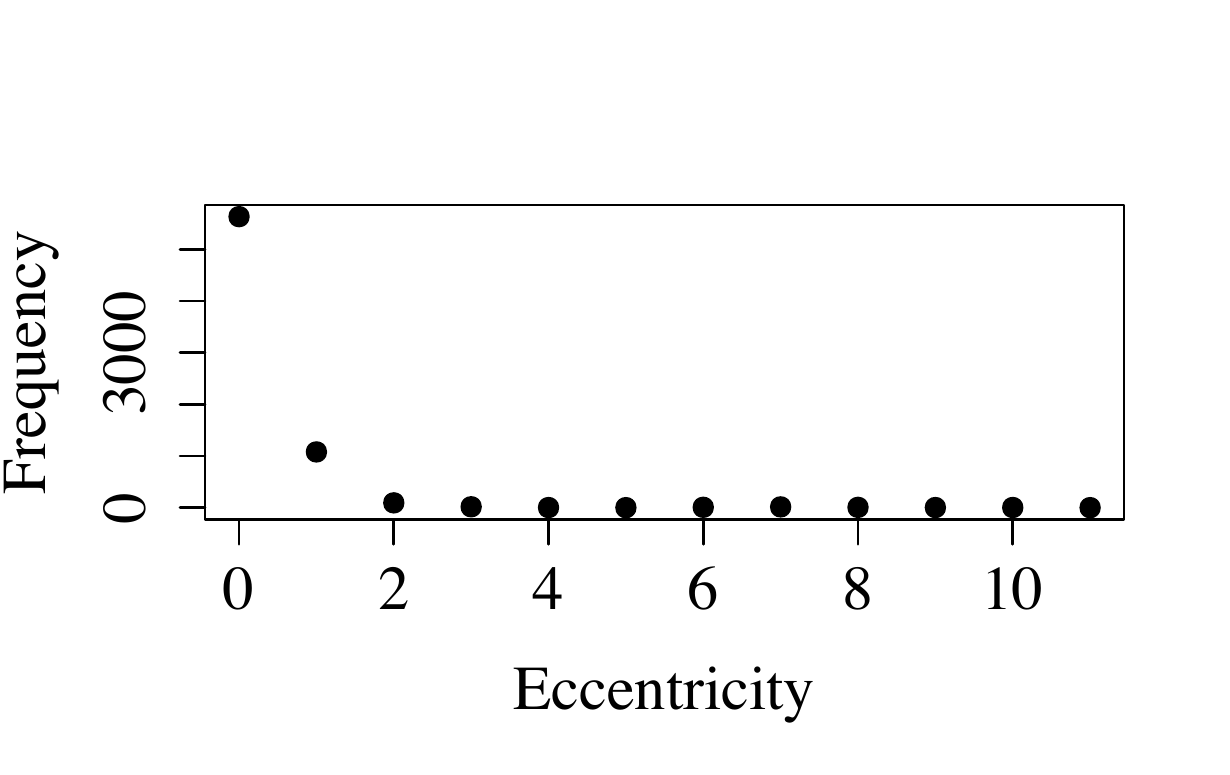}
\caption{Frequency Distribution of Eccentricity}
\label{fig:eccenHist}
\end{figure}

The distribution of eccentricity in Figure~\ref{fig:netPar} indicates that the probability of values greater than five is less than 0.5 and maximum eccentricity has a probability close to zero. It reveals that there are a few resources which have great proximity to others. Manual investigation shows HiddenWiki, onionDir, TorWiki, and Runion Dark Forum have eccentricities near the maximum. 
%
To get a deeper insight into the distribution of eccentricity values, Figure~\ref{fig:eccenHist} shows the eccentricity histogram where for 5,635 nodes (82\% of all vertices), this measure is zero. As the network has several connected components, the distance for nodes located in different components is defined as infinite which makes their corresponding eccentricity tend towards zero. Also, since there is no outgoing edge from surface websites to other nodes in the network, their eccentricity will be zero. Such surface websites account for 64\% of all vertices with zero eccentricity. The rest (36\% of zero centralities and 61\% of Tor domains) belong to dark services which are either isolated or have only a few connections with others.
On the other hand, for 1,102 nodes (16\% of all vertices and 33\% of Tor domains), the centrality value is in $\left[1,2\right)$ and a few (2\% of all and 6\% of Tor domains) have eccentricity equal to or greater than 2. This finding reveals in spite of referencing to surface, a larger portion of Tor domains are far from either dark or surface resources which 
complements the analysis in~\cite{zabihimayvan2019broad} that reports lack of tendency in Tor domains to link each other.  

\textbf{3. Stress centrality.} The stress centrality shows the number of shortest paths running through each vertex in a network with no multiple edges~\cite{brandes2001faster}. Higher value of stress for a dark domain or surface website shows it is traversed by a large number of shortest paths and thus plays an important role in information dissemination trough the network. However, the corresponding vertex may not be critical in maintaining the communication between pairs of nodes since there may be other shortest paths between each pair that are not passed over it. Let $\sigma_{st}(v)$ show the number of shortest path between two nodes of $s$ and $t$ that contain vertex $v$. Then, the stress centrality of $v$, $Cent_{s}(v)$, is calculated as follows:
$$
Cent_{s}(v)=\sum_{s\neq t\neq v \in V}\sigma_{st}(v)
$$
which results in the summation of partial centralities for all pairs of nodes in the network.
\begin{table}
\caption{20-top Tor Services with high Stress centrality}
\centering
\begin{tabular} { c | l | c | l }
\toprule
Service address & Service type/name & Service address & Service type/name \\
\midrule
24boths2mh6sxaz5 & Automated bots for rent (Russian) & kpvz7ki2lzvnwve7 & HiddenWiki\\
\hline
24xbtc424rgg5zah & Digital Crypto (Russian) & hosting6iar5zo7c & Real Hosting\\
\hline
2qrdpvonwwqnic7j & Dark net Community (Italian) & gnvweaoe2xzjqldu & VisiTor\\
\hline
32pbf32xi6ccm63z & Pornography & 2ddjd7xsni7pefcx & Fresh Onions\\
\hline
3bbaaaccczcbdddz & OnionLand Search Engine & darkwikiss2rmeby & DarkWiki (Russian)\\
\hline
3kyl4i7bfdgwelmf & DeepDotWeb & deurfnquin7mvni2 & Pyrowiki|Pyrotechnics drug wiki\\
\hline
doe6ypf2fcyznaq5 & Runion Forum (Russian) & torlinkbgs6aabns & HiddenWiki\\
\hline
dirnxxdraygbifgc & OnionDir & wiki5kauuihowqi5 & HiddenWiki\\
\hline
4do6yq4iwstidagh & Debian Onion Service & newsiiwanaduqpre & News blog on Tor\\
\hline
4hohkxjvlt5fjzqv & Search Engine (Portuguese) & lwplxqzvmgu43uff & Runion Forum (Russian)\\
\bottomrule
\end{tabular}
\label{tab:strsExamp}
\end{table}
According to the Table~\ref{tab:params}, the average stress centrality of the network is 175 while the maximum equals to 230,857. As Figure~\ref{fig:netPar} indicates, the probability of values greater than the average is rapidly decreasing from 0.5 while the probability for the rest is approximately 0.5. Investigation depicts that most network vertices (94\%) have stress lower than the average and only a few centralities (6\%) are greater than this value. Also, since surface websites have no out-going edges to others, their stress centrality should be zero. Therefore, the minor large centralities belong to Tor services. This implies that a large portion of Tor services are reluctant to contribute in information traversing through the network which help them remain difficult to discover by Tor users, especially by the law enforcement agencies.   

Assuming Tor domains sorted in descending order based on their stress centrality, Table~\ref{tab:strsExamp} illustrates the 20-top dark services which provide information in different languages including Russian and Portuguese. More than half of these services belong to well-known directories and their high stress centrality is the result of their huge number of in-coming and out-going connections. There are also two forums, one Bitcoin, and one bot rental domain that provide services in Russian. 
There are also one pornography, one news blog, and one hosting service which show the other main Tor domains conductive to information transmission through the dark Web.

\textbf{4. Out-degree.} The out-degree is another important feature which gives the number of links each domain or website has to others. According to Table~\ref{tab:params}, values range from 0 to 1,358 where the average is 4.29. The distribution of values in Figure~\ref{fig:netPar} indicates that the probability of values less than the average is approximately 1. Most vertices (81\%) have no link to others which includes all surface websites along with 60\% of Tor domains. 
This implies that more than half of Tor services are reluctant to refer their users to either surface or dark sources. For the remaining 40\%, their out-degree lies in $\left[1, 1358\right]$ and based on the manual investigation, only 14 domains have out-degree centralities greater than 400.  
\begin{table}
\caption{20-top Tor Services without-degree centrality greater than 400}
\centering
\begin{tabular} { c | p{40mm} | c | l }
\toprule
Domain name & Service type & Domain name & Service type \\
\midrule
dirnxxdraygbifgc & Dir (OnionDir) & jh32yv5zgayyyts3 & Dir (Onion List)\\
\hline
4sy6ebszykvcv2n6 & News (German) & wikitjerrta4qgz4 & Dir (HiddenWiki)\\
\hline
4zkgktc6hjdmp6kw & Dir (HiddenWiki) & chws5ibwliag4fyc & Dir (OnionLand Search)\\
\hline
5jgis47vdcpaeafp & Dir (La Wiki Oculta)/Spanish version of HiddenWiki & wiki5kauuihowqi5 & Dir (HiddenWiki)\\
\hline
gnvweaoe2xzjqldu & Multimedia (hosting Debian onion services) & hiddenwik55b36km & Dir (HiddenWiki)
\\
\hline
7g5bqm7htspqauum & Dir (HiddenWiki) & torwikignoueupfm & Dir (TorWiki)\\
\hline
2ddjd7xsni7pefcx & Dir (Fresh Onions) & zgrl6sghf5jh37zz & Dir (HiddenWiki)
\\
\bottomrule
\end{tabular}
\label{tab:outdExamp}
\end{table}
Table~\ref{tab:outdExamp} presents the description of services provided by the domains with more than 400 links to others. As illustrated, it contains popular directories, one German news, and one multimedia service that provide information on hosting Debian onion services. This finding is not surprising as directories are the services with huge lists of onion addresses in forms of hyperlinks that results in high out-degree centrality. News domains also provide information in different topics and may support the validity of the content by linking to the original websites of the news. Manual investigation of the content provided by the multimedia service also depicts that this domain contains several links to other similar resources on both Tor and surface Web which provide technical details about Tor and Debian operating system. 

\textbf{5. In-degree.} This feature demonstrates the number of links to a dark domain or surface website. As Table~\ref{tab:params} indicates, values range between zero and 309 with the average of 4.29. Probability of values less than the average is close to 0.5 while probability of values more than 100 is less than 0.1. Manual exploring reveals that only 8.8\% of vertices have less-than-average in-degree, 91\% have in-degree values between the average and 100, and only 10 nodes have more than 100 in-coming links. To further investigate such outliers, Table~\ref{tab:indExamp} lists their address/names and whether each is a dark or surface resource. 
\begin{table}
\caption{10-top Tor Services with In-degree centrality greater than 100}
\centering
\begin{tabular} { p{40mm} | l }
\toprule
Node Web address & Node type \\
\midrule
blockchainbdgpzk.onion (along with its 3 mirrors) & Dark Bitcoin service\\
\hline
btc.com & Surface Bitcoin Block Explorer\\
\hline
twitter.com & Surface Twitter website\\
\hline
Facebook.com & Surface Facebook website\\
\hline
torproject.org & Surface Tor project website\\
\hline
github.com & Surface GitHub website\\
\hline
3g2upl4pq6kufc4m.onion & Dark DuckDuckGo Search Engine\\
\bottomrule
\end{tabular}
\label{tab:indExamp}
\end{table}
As Table~\ref{tab:indExamp} represents, one dark and surface bitcoin services are among the nodes with more than 100 in-coming links. It is not surprising because there are marketplaces on dark Web which use digital cryptocurrency using either dark or surface Web services as payment method. The \textit{DuckDuckGo} search engine is another dark service that has notable in-degree value. It shows that there are several Tor domains which refer their visitors to this search engine for further exploration on the dark Web.
There are also some popular surface websites such as Twitter, Facebook, GitHub, and Tor project which are linked by several Tor services which indicates their popularity among dark services.  

\textbf{6. Clustering Coefficient (local).} The local clustering coefficient evaluates how much neighbors of a node tend to cluster with each other. In the context of dark services, this metric for a service $c$ indicates the extent to which other services that are directly linked by $c$ make reference to each other. 
This metric calculates the fraction of number of edges among all neighbors of a node over maximum number of possible edges its neighbors can have with each other. Assuming $\eta_v$ is the number of neighbors of vertex $v$ and $e_v$ is the number of connected pairs between all neighbors of $v$~\cite{barabasi2004network}, the local clustering coefficient of $v$ in a directed network is calculated as: 
$$
CC_v = \frac{e_v}{\eta_v(\eta_v-1)}
$$

Ranging between 0 to 1, values close or equal to zero indicate a node whose neighbors seldom connect to each other while values near one show a node with tightly connected neighborhood. 

As Tabel~\ref{tab:params} and Figure~\ref{fig:netPar} indicate, the average clustering coefficient is 0.09 and the probability of values less than the average is near 0.5 which indicates that for about half of the vertices, neighborhoods are almost sparse and barely link each other. As mentioned before, surface websites in the network have no link to other nodes and hence, their clustering coefficient will be zero. Regardless of the surface nodes, the rest with less-than-average clustering coefficient forms 61\% of the Tor domains while the remaining 39\% have values greater than 0.09. 
This finding implies that in spite of interest in Tor domains to remain isolated, references to surface websites increase the clustering coefficient values for dark services. 
Also, the reason why their clustering coefficients have notable difference with 1 (as the maximum possible value) is that there is no links from surface websites to others, which reduces the number of links between pairs of nodes in neighborhoods of vertices.

\textbf{7. Average Shortest Path Length.} This feature is one of the most important characteristics of real-world complex networks~\cite{mao2013analysis}. This metric indicates the expected distance between each connected pairs of nodes. Assuming $G=(V,E)$ is a directed graph, $dist(s,t)$ is defined as the shortest distance from $s$ to $t$, and $\alpha_G$ shows the number of all paths in $G$, $\overline {SP}(G)$ which indicates the average shortest path length of $G$ is calculated as:
$$
\overline {SP}(G)=\frac{\sum_{v_i,v_j \in V}^{|V|} dist(v_i,v_j)}{\alpha_G}
$$
where for an isolated vertex, the distance to any other node is zero. Similarly, if $v_i=v_j$, the average shortest path length will be zero. 

Based on Table~\ref{tab:params}, the mean (.2) and maximum (6.64) values of this parameter demonstrate that every node of the network can be reached from others using by average one and at most seven edge(s), respectively.
As Figure~\ref{fig:netPar} indicates, the probability of values less than or equal to 2 is close to 1 while for values near 7, the probability is less than 0.2. This implies that most vertices are accessible via at most two edges and only a few of them are far from the others. As mentioned earlier, the linking process of the network indicates that large portion of vertices participate in a single large connected component which can explain why the average shortest path length of most vertices have low values.
Since the network is almost connected and the maximum average shortest path is comparable with the logarithm of network size, users of this small world can reach every dark domain or surface website after traversing a short distance on the network.   
\section{Conclusion and Future work}
\label{sec:con}

This work investigates the network of references from dark services to surface websites using well-known network analysis metrics to answer how the network structure of dark-to-surface hyperlinks differs with the linking pattern of Tor hidden services. It also considers to what extent dark services are vulnerable against information leakage caused by linking to the surface Web resources. To achieve a deeper insight into the nature of Tor hidden services, the results are also analyzed regarding the type of content and information provided by Tor domains. 

Results depict a structure that contrasts with the hyperlink structure of Tor. The dark-to-surface reference network is a single massive connected component with small number of isolated Tor domains. However, Tor resources are still interested in being isolated.
Only a few Tor hidden services are immune against information leakage while over 90\% have at least one link to the surface Web. 
Such referencing to surface makes Tor domains closer to each other and increases their interest to cluster. But, it does not raise the contribution of dark services to either communication or information dissemination through the network. Analyses on the type of dark services reveal that well-known Tor directories have the maximum proximity to all the other Web resources and significantly contribute to both communication and information dissemination from the dark-to-surface network. A few domains with other types of services such as news, pornography, and multimedia also contribute to information propagation through the network. Analysis on the degree distribution indicates 
popular surface websites such as GitHub, Twitter, and Facebook are popular even among Tor domains.

Future work can expand this work further by investigating the changes in the pattern of referencing from dark to surface Web over time. It can reveal some permanent characteristics in the linking structure of dark services and indicate how their vulnerability against information leakage can change over time. Another direction of future work is to compare this network with other type of socio-technical systems like the Web structure. Such a comparative study can reveal possible ways to model the dark-to-surface network using algorithms such as ERGM which can be useful for link prediction and thus, assessing the risk of information leakage.

\section{Acknowledgments}

This paper is based on work supported by the National Science Foundation (NSF) under Grant No. 1464104. Any opinions, findings, and conclusions or recommendations expressed are those of the author(s) and do not necessarily reflect the views of the NSF.

\bibliographystyle{ACM-Reference-Format}
\bibliography{sample-base}


\begin{thebibliography}{29}


\ifx \showCODEN    \undefined \def \showCODEN     #1{\unskip}     \fi
\ifx \showDOI      \undefined \def \showDOI       #1{#1}\fi
\ifx \showISBNx    \undefined \def \showISBNx     #1{\unskip}     \fi
\ifx \showISBNxiii \undefined \def \showISBNxiii  #1{\unskip}     \fi
\ifx \showISSN     \undefined \def \showISSN      #1{\unskip}     \fi
\ifx \showLCCN     \undefined \def \showLCCN      #1{\unskip}     \fi
\ifx \shownote     \undefined \def \shownote      #1{#1}          \fi
\ifx \showarticletitle \undefined \def \showarticletitle #1{#1}   \fi
\ifx \showURL      \undefined \def \showURL       {\relax}        \fi
\providecommand\bibfield[2]{#2}
\providecommand\bibinfo[2]{#2}
\providecommand\natexlab[1]{#1}
\providecommand\showeprint[2][]{arXiv:#2}

\bibitem[\protect\citeauthoryear{Barabasi and Oltvai}{Barabasi and
  Oltvai}{2004}]%
        {barabasi2004network}
\bibfield{author}{\bibinfo{person}{Albert-Laszlo Barabasi} {and}
  \bibinfo{person}{Zoltan~N Oltvai}.} \bibinfo{year}{2004}\natexlab{}.
\newblock \showarticletitle{Network biology: understanding the cell's
  functional organization}.
\newblock \bibinfo{journal}{\emph{Nature reviews genetics}}
  \bibinfo{volume}{5}, \bibinfo{number}{2} (\bibinfo{year}{2004}),
  \bibinfo{pages}{101}.
\newblock


\bibitem[\protect\citeauthoryear{Bauer, McCoy, Grunwald, Kohno, and
  Sicker}{Bauer et~al\mbox{.}}{2007}]%
        {bauer2007low}
\bibfield{author}{\bibinfo{person}{Kevin Bauer}, \bibinfo{person}{Damon McCoy},
  \bibinfo{person}{Dirk Grunwald}, \bibinfo{person}{Tadayoshi Kohno}, {and}
  \bibinfo{person}{Douglas Sicker}.} \bibinfo{year}{2007}\natexlab{}.
\newblock \showarticletitle{Low-resource routing attacks against tor}. In
  \bibinfo{booktitle}{\emph{Proc. of the 2007 ACM workshop on Privacy in
  electronic society}}. \bibinfo{pages}{11--20}.
\newblock


\bibitem[\protect\citeauthoryear{Bernaschi, Celestini, Guarino, and
  Lombardi}{Bernaschi et~al\mbox{.}}{2017}]%
        {bernaschi2017exploring}
\bibfield{author}{\bibinfo{person}{Massimo Bernaschi},
  \bibinfo{person}{Alessandro Celestini}, \bibinfo{person}{Stefano Guarino},
  {and} \bibinfo{person}{Flavio Lombardi}.} \bibinfo{year}{2017}\natexlab{}.
\newblock \showarticletitle{Exploring and analyzing the Tor hidden services
  graph}.
\newblock \bibinfo{journal}{\emph{ACM Transactions on the Web}}
  \bibinfo{volume}{11}, \bibinfo{number}{4} (\bibinfo{year}{2017}),
  \bibinfo{pages}{24}.
\newblock


\bibitem[\protect\citeauthoryear{Bernaschi, Celestini, Guarino, Lombardi, and
  Mastrostefano}{Bernaschi et~al\mbox{.}}{2019}]%
        {bernaschi2019spiders}
\bibfield{author}{\bibinfo{person}{Massimo Bernaschi},
  \bibinfo{person}{Alessandro Celestini}, \bibinfo{person}{Stefano Guarino},
  \bibinfo{person}{Flavio Lombardi}, {and} \bibinfo{person}{Enrico
  Mastrostefano}.} \bibinfo{year}{2019}\natexlab{}.
\newblock \showarticletitle{Spiders like Onions: on the Network of Tor Hidden
  Services}. In \bibinfo{booktitle}{\emph{The World Wide Web Conference}}.
  \bibinfo{pages}{105--115}.
\newblock


\bibitem[\protect\citeauthoryear{Biryukov and Pustogarov}{Biryukov and
  Pustogarov}{2015}]%
        {biryukov2015bitcoin}
\bibfield{author}{\bibinfo{person}{Alex Biryukov} {and} \bibinfo{person}{Ivan
  Pustogarov}.} \bibinfo{year}{2015}\natexlab{}.
\newblock \showarticletitle{Bitcoin over Tor isn't a Good Idea}. In
  \bibinfo{booktitle}{\emph{2015 IEEE Symposium on Security and Privacy}}.
  \bibinfo{pages}{122--134}.
\newblock


\bibitem[\protect\citeauthoryear{Biryukov, Pustogarov, and Weinmann}{Biryukov
  et~al\mbox{.}}{2013}]%
        {trawling}
\bibfield{author}{\bibinfo{person}{Alex Biryukov}, \bibinfo{person}{Ivan
  Pustogarov}, {and} \bibinfo{person}{Ralf-Philipp Weinmann}.}
  \bibinfo{year}{2013}\natexlab{}.
\newblock \showarticletitle{Trawling for Tor Hidden Services: Detection,
  measurement, deanonymization}. In \bibinfo{booktitle}{\emph{Symposium on
  Security and Privacy}}. \bibinfo{pages}{80--94}.
\newblock


\bibitem[\protect\citeauthoryear{Blei, Ng, and Jordan}{Blei
  et~al\mbox{.}}{2003}]%
        {lda}
\bibfield{author}{\bibinfo{person}{David~M Blei}, \bibinfo{person}{Andrew~Y
  Ng}, {and} \bibinfo{person}{Michael~I Jordan}.}
  \bibinfo{year}{2003}\natexlab{}.
\newblock \showarticletitle{Latent Dirichlet Allocation}.
\newblock \bibinfo{journal}{\emph{Journal of Machine Learning Research}}
  \bibinfo{volume}{3}, \bibinfo{number}{Jan} (\bibinfo{year}{2003}),
  \bibinfo{pages}{993--1022}.
\newblock


\bibitem[\protect\citeauthoryear{Brandes}{Brandes}{2001}]%
        {brandes2001faster}
\bibfield{author}{\bibinfo{person}{Ulrik Brandes}.}
  \bibinfo{year}{2001}\natexlab{}.
\newblock \showarticletitle{A faster algorithm for betweenness centrality}.
\newblock \bibinfo{journal}{\emph{Journal of mathematical sociology}}
  \bibinfo{volume}{25}, \bibinfo{number}{2} (\bibinfo{year}{2001}),
  \bibinfo{pages}{163--177}.
\newblock


\bibitem[\protect\citeauthoryear{Broder, Kumar, Maghoul, Raghavan, Rajagopalan,
  Stata, Tomkins, and Wiener}{Broder et~al\mbox{.}}{2000}]%
        {connectedWeb}
\bibfield{author}{\bibinfo{person}{Andrei Broder}, \bibinfo{person}{Ravi
  Kumar}, \bibinfo{person}{Farzin Maghoul}, \bibinfo{person}{Prabhakar
  Raghavan}, \bibinfo{person}{Sridhar Rajagopalan}, \bibinfo{person}{Raymie
  Stata}, \bibinfo{person}{Andrew Tomkins}, {and} \bibinfo{person}{Janet
  Wiener}.} \bibinfo{year}{2000}\natexlab{}.
\newblock \showarticletitle{Graph Structure in the Web}.
\newblock \bibinfo{journal}{\emph{Computer Networks}} \bibinfo{volume}{33},
  \bibinfo{number}{1-6} (\bibinfo{year}{2000}), \bibinfo{pages}{309--320}.
\newblock


\bibitem[\protect\citeauthoryear{Burda, Boot, and Allodi}{Burda
  et~al\mbox{.}}{2019}]%
        {burda2019characterizing}
\bibfield{author}{\bibinfo{person}{Pavlo Burda}, \bibinfo{person}{Coen Boot},
  {and} \bibinfo{person}{Luca Allodi}.} \bibinfo{year}{2019}\natexlab{}.
\newblock \showarticletitle{Characterizing the redundancy of DarkWeb. onion
  services}. In \bibinfo{booktitle}{\emph{Proc. of the 14th Int. Conference on
  Availability, Reliability and Security}}. \bibinfo{pages}{19}.
\newblock


\bibitem[\protect\citeauthoryear{Cambiaso, Vaccari, Patti, and Aiello}{Cambiaso
  et~al\mbox{.}}{2019}]%
        {cambiaso2019darknet}
\bibfield{author}{\bibinfo{person}{Enrico Cambiaso}, \bibinfo{person}{Ivan
  Vaccari}, \bibinfo{person}{Luca Patti}, {and} \bibinfo{person}{Maurizio
  Aiello}.} \bibinfo{year}{2019}\natexlab{}.
\newblock \showarticletitle{Darknet Security: A Categorization of Attacks to
  the Tor Network}. In \bibinfo{booktitle}{\emph{Italian Conference on
  Cybersecurity}}.
\newblock


\bibitem[\protect\citeauthoryear{Chertoff and Simon}{Chertoff and
  Simon}{2015}]%
        {cyberSecurity}
\bibfield{author}{\bibinfo{person}{Michael Chertoff} {and}
  \bibinfo{person}{Tobby Simon}.} \bibinfo{year}{2015}\natexlab{}.
\newblock \showarticletitle{The Impact of the Dark Web on Internet Governance
  and Cyber Security}.
\newblock \bibinfo{journal}{\emph{GCIG Paper Series}} \bibinfo{number}{6}
  (\bibinfo{year}{2015}).
\newblock


\bibitem[\protect\citeauthoryear{Clarke, Sandberg, Toseland, and
  Verendel}{Clarke et~al\mbox{.}}{2010}]%
        {freenet}
\bibfield{author}{\bibinfo{person}{Ian Clarke}, \bibinfo{person}{Oskar
  Sandberg}, \bibinfo{person}{Matthew Toseland}, {and} \bibinfo{person}{Vilhelm
  Verendel}.} \bibinfo{year}{2010}\natexlab{}.
\newblock \showarticletitle{Private Communication through a Network of Trusted
  Connections: The Dark Freenet}.
\newblock \bibinfo{journal}{\emph{Network}} (\bibinfo{year}{2010}).
\newblock


\bibitem[\protect\citeauthoryear{Goldschlag, Reed, and Syverson}{Goldschlag
  et~al\mbox{.}}{1999}]%
        {goldschlag1999onion}
\bibfield{author}{\bibinfo{person}{David Goldschlag}, \bibinfo{person}{Michael
  Reed}, {and} \bibinfo{person}{Paul Syverson}.}
  \bibinfo{year}{1999}\natexlab{}.
\newblock \bibinfo{booktitle}{\emph{Onion routing for anonymous and private
  internet connections}}.
\newblock \bibinfo{type}{{T}echnical {R}eport}. \bibinfo{institution}{NAVAL
  Research Lab Washington DC Center For High Assurance Computing Ssystems}.
\newblock


\bibitem[\protect\citeauthoryear{Griffith, Xu, and Ratti}{Griffith
  et~al\mbox{.}}{2017}]%
        {griffith2017graph}
\bibfield{author}{\bibinfo{person}{Virgil Griffith}, \bibinfo{person}{Yang Xu},
  {and} \bibinfo{person}{Carlo Ratti}.} \bibinfo{year}{2017}\natexlab{}.
\newblock \showarticletitle{Graph Theoretic Properties of the Darkweb}.
\newblock \bibinfo{journal}{\emph{arXiv preprint arXiv:1704.07525}}
  (\bibinfo{year}{2017}).
\newblock


\bibitem[\protect\citeauthoryear{Hage and Harary}{Hage and Harary}{1995}]%
        {hage1995eccentricity}
\bibfield{author}{\bibinfo{person}{Per Hage} {and} \bibinfo{person}{Frank
  Harary}.} \bibinfo{year}{1995}\natexlab{}.
\newblock \showarticletitle{Eccentricity and centrality in networks}.
\newblock \bibinfo{journal}{\emph{Social networks}} \bibinfo{volume}{17},
  \bibinfo{number}{1} (\bibinfo{year}{1995}), \bibinfo{pages}{57--63}.
\newblock


\bibitem[\protect\citeauthoryear{Henri}{Henri}{2017}]%
        {riffle}
\bibfield{author}{\bibinfo{person}{Vanessa Henri}.}
  \bibinfo{year}{2017}\natexlab{}.
\newblock \showarticletitle{The Dark Web: Some Thoughts for an Educated
  Debate}.
\newblock \bibinfo{journal}{\emph{Canadian Journal of Law and Technology}}
  \bibinfo{volume}{15}, \bibinfo{number}{1} (\bibinfo{year}{2017}).
\newblock


\bibitem[\protect\citeauthoryear{Hulpus, Hayes, Karnstedt, and Greene}{Hulpus
  et~al\mbox{.}}{2013}]%
        {gbtl}
\bibfield{author}{\bibinfo{person}{Ioana Hulpus}, \bibinfo{person}{Conor
  Hayes}, \bibinfo{person}{Marcel Karnstedt}, {and} \bibinfo{person}{Derek
  Greene}.} \bibinfo{year}{2013}\natexlab{}.
\newblock \showarticletitle{Unsupervised Graph-based Topic Labelling using
  DBpedia}. In \bibinfo{booktitle}{\emph{Proc. of the sixth ACM Int. Conf. on
  Web Search and Data Mining}}. \bibinfo{pages}{465--474}.
\newblock


\bibitem[\protect\citeauthoryear{Mao and Zhang}{Mao and Zhang}{2013}]%
        {mao2013analysis}
\bibfield{author}{\bibinfo{person}{Guoyong Mao} {and} \bibinfo{person}{Ning
  Zhang}.} \bibinfo{year}{2013}\natexlab{}.
\newblock \showarticletitle{Analysis of average shortest-path length of
  scale-free network}.
\newblock \bibinfo{journal}{\emph{Journal of Applied Mathematics}}
  \bibinfo{volume}{2013} (\bibinfo{year}{2013}).
\newblock


\bibitem[\protect\citeauthoryear{McCoy, Bauer, Grunwald, Kohno, and
  Sicker}{McCoy et~al\mbox{.}}{2008}]%
        {shining}
\bibfield{author}{\bibinfo{person}{Damon McCoy}, \bibinfo{person}{Kevin Bauer},
  \bibinfo{person}{Dirk Grunwald}, \bibinfo{person}{Tadayoshi Kohno}, {and}
  \bibinfo{person}{Douglas Sicker}.} \bibinfo{year}{2008}\natexlab{}.
\newblock \showarticletitle{Shining Light in Dark Places: Understanding the Tor
  Network}. In \bibinfo{booktitle}{\emph{Privacy Enhancing Technologies}}.
  \bibinfo{pages}{63--76}.
\newblock
\showISBNx{978-3-540-70630-4}


\bibitem[\protect\citeauthoryear{Mohaisen and Ren}{Mohaisen and Ren}{2017}]%
        {leakage}
\bibfield{author}{\bibinfo{person}{Aziz Mohaisen} {and} \bibinfo{person}{Kui
  Ren}.} \bibinfo{year}{2017}\natexlab{}.
\newblock \showarticletitle{Leakage of. onion at the DNS Root: Measurements,
  Causes, and Countermeasures}.
\newblock \bibinfo{journal}{\emph{IEEE/ACM Transactions on Networking}}
  \bibinfo{volume}{25}, \bibinfo{number}{5} (\bibinfo{year}{2017}),
  \bibinfo{pages}{3059--3072}.
\newblock


\bibitem[\protect\citeauthoryear{Newman and Girvan}{Newman and Girvan}{2004}]%
        {newman2004finding}
\bibfield{author}{\bibinfo{person}{Mark~EJ Newman} {and}
  \bibinfo{person}{Michelle Girvan}.} \bibinfo{year}{2004}\natexlab{}.
\newblock \showarticletitle{Finding and evaluating community structure in
  networks}.
\newblock \bibinfo{journal}{\emph{Physical review E}} \bibinfo{volume}{69},
  \bibinfo{number}{2} (\bibinfo{year}{2004}), \bibinfo{pages}{026113}.
\newblock


\bibitem[\protect\citeauthoryear{Sanatinia and Noubir}{Sanatinia and
  Noubir}{2017}]%
        {sanatinia2017off}
\bibfield{author}{\bibinfo{person}{Amirali Sanatinia} {and}
  \bibinfo{person}{Guevara Noubir}.} \bibinfo{year}{2017}\natexlab{}.
\newblock \showarticletitle{Off-path man-in-the-middle attack on tor hidden
  services}.
\newblock \bibinfo{journal}{\emph{New England Security Day}}
  (\bibinfo{year}{2017}).
\newblock


\bibitem[\protect\citeauthoryear{Sanatinia, Park, Blass, Mohaisen, and
  Noubir}{Sanatinia et~al\mbox{.}}{2019}]%
        {sanatinia2019privacy}
\bibfield{author}{\bibinfo{person}{Amirali Sanatinia}, \bibinfo{person}{Jeman
  Park}, \bibinfo{person}{Erik-Oliver Blass}, \bibinfo{person}{Aziz Mohaisen},
  {and} \bibinfo{person}{Guevara Noubir}.} \bibinfo{year}{2019}\natexlab{}.
\newblock \showarticletitle{A Privacy-Preserving Longevity Study of Tor's
  Hidden Services}.
\newblock \bibinfo{journal}{\emph{arXiv preprint arXiv:1909.03576}}
  (\bibinfo{year}{2019}).
\newblock


\bibitem[\protect\citeauthoryear{Sanchez-Rola, Balzarotti, and
  Santos}{Sanchez-Rola et~al\mbox{.}}{2017}]%
        {onionEyes}
\bibfield{author}{\bibinfo{person}{Iskander Sanchez-Rola},
  \bibinfo{person}{Davide Balzarotti}, {and} \bibinfo{person}{Igor Santos}.}
  \bibinfo{year}{2017}\natexlab{}.
\newblock \showarticletitle{The Onions Have Eyes: A Comprehensive Structure and
  Privacy Analysis of Tor Hidden Services}. In \bibinfo{booktitle}{\emph{Proc.
  of the 26th Int. Conf. on World Wide Web}}. \bibinfo{pages}{1251--1260}.
\newblock
\showISBNx{978-1-4503-4913-0}


\bibitem[\protect\citeauthoryear{Syverson, Dingledine, and Mathewson}{Syverson
  et~al\mbox{.}}{2004}]%
        {secondGeneration}
\bibfield{author}{\bibinfo{person}{Paul Syverson}, \bibinfo{person}{R
  Dingledine}, {and} \bibinfo{person}{N Mathewson}.}
  \bibinfo{year}{2004}\natexlab{}.
\newblock \showarticletitle{Tor: The Second-generation Onion Router}. In
  \bibinfo{booktitle}{\emph{Usenix Security}}.
\newblock


\bibitem[\protect\citeauthoryear{Xu and Chen}{Xu and Chen}{2008}]%
        {topology}
\bibfield{author}{\bibinfo{person}{Jennifer Xu} {and} \bibinfo{person}{Hsinchun
  Chen}.} \bibinfo{year}{2008}\natexlab{}.
\newblock \showarticletitle{The topology of Dark Networks}.
\newblock \bibinfo{journal}{\emph{Commun. ACM}} \bibinfo{volume}{51},
  \bibinfo{number}{10} (\bibinfo{year}{2008}), \bibinfo{pages}{58--65}.
\newblock


\bibitem[\protect\citeauthoryear{Zabihimayvan and Doran}{Zabihimayvan and
  Doran}{2018}]%
        {ccdf}
\bibfield{author}{\bibinfo{person}{Mahdieh Zabihimayvan} {and}
  \bibinfo{person}{Derek Doran}.} \bibinfo{year}{2018}\natexlab{}.
\newblock \showarticletitle{Some (Non-) Universal Features of Web Robot
  Traffic}. In \bibinfo{booktitle}{\emph{Annual Conf. on Information Sciences
  and Systems}}. \bibinfo{pages}{1--6}.
\newblock


\bibitem[\protect\citeauthoryear{Zabihimayvan, Sadeghi, Doran, and
  Allahyari}{Zabihimayvan et~al\mbox{.}}{2019}]%
        {zabihimayvan2019broad}
\bibfield{author}{\bibinfo{person}{Mahdieh Zabihimayvan}, \bibinfo{person}{Reza
  Sadeghi}, \bibinfo{person}{Derek Doran}, {and} \bibinfo{person}{Mehdi
  Allahyari}.} \bibinfo{year}{2019}\natexlab{}.
\newblock \showarticletitle{A Broad Evaluation of the Tor English Content
  Ecosystem}. In \bibinfo{booktitle}{\emph{Proc. of the 10th ACM Conference on
  Web Science}}. \bibinfo{pages}{333--342}.
\newblock


\end{thebibliography}

\end{document}